\documentclass{osa-article}
\journal{oe}

\usepackage[utf8]{inputenc}
\usepackage{amsmath}
\usepackage{graphicx}
\usepackage{subcaption}
\usepackage{datetime}
\usepackage{comment}

\begin{document}
\title{Dark Current and Single Photon Detection by 1550 nm Avalanche Photodiodes: Dead Time Corrected Probability Distributions and Entropy Rates }

\author{Nicole Menkart,\authormark{1,2,*} Joseph D. Hart,\authormark{3} Thomas E. Murphy,\authormark{1,2}, \& Rajarshi Roy \authormark{2,4}}

\address{\authormark{1}Department of Electrical \& Computer Engineering, University of Maryland, College Park, College Park, MD 20742, USA\\
\authormark{2}Institute for Research in Electronics and Applied Physics, University of Maryland, College Park, College Park, MD 20742, USA\\
\authormark{3}Optical Sciences Division, U.S. Naval Research Laboratory, Washington, DC, 20375, USA\\
\authormark{4}Department of Physics, University of Maryland, College Park, College Park, MD 20742, USA}

\email{\authormark{*}nmenkart@umd.edu}

\begin{abstract}
Single photon detectors have dark count rates that depend strongly on the bias level for detector operation. In the case of weak light sources such as novel lasers or single-photon emitters, the rate of counts due to the light source can be comparable to that of the detector dark counts. In such cases, a characterization of the statistical properties of the dark counts is necessary. The dark counts are often assumed to follow a Poisson process that is statistically independent of the incident photon counts. This assumption must be validated for specific types of photodetectors. In this work, we focus on single-photon avalanche photodiodes (SPADs) made for 1550nm. For the InGaAs detectors used, we find the measured distributions often differ significantly from Poisson due to the presence of dead time and afterpulsing with the difference increasing with the bias level used for obtaining higher quantum efficiencies. We find that when the dead time is increased to remove the effects of afterpulsing, it is necessary to correct the measured distributions for the effects of the dead time. To this end, we apply an iterative algorithm to remove dead time effects from the probability distribution for dark counts as well as for the case where light from an external weak laser source (known to be Poisson) is detected together with the dark counts. We believe this to be the first instance of the comprehensive application of this algorithm to real data and find that the dead time corrected probability distributions are Poisson distributions in both cases. We additionally use the Grassberger-Procaccia algorithm to estimate the entropy production rates of the dark count processes, which provides a single metric that characterizes the temporal correlations between dark counts as well as the shape of the distribution. We have thus developed a systematic procedure for taking data with 1550nm SPADs and obtaining accurate photocount statistics to examine novel light sources.
\end{abstract}

\section{Introduction}
Single photon counting has been used for many decades to study the quantum properties of light and its interactions with atoms and molecules. Single photon detection was used to lay the foundations of coherent states emitted by laser sources and non-classical states of light fields \cite{mandel_optical_1995,loudon_quantum_2000}. More recently, it is an indispensable technique for the characterization of single photon emitters (SPEs) which are used widely in quantum information applications \cite{nielsen_quantum}. Traditionally, photomultiplier tubes were used to count photons \cite{saleh_photoelectron_1978} in the visible and ultraviolet spectrum, but with the development of fiber optic telecommunications, semiconductor-based detectors in the 1550nm region came into widespread use for weak signals at the single photon level. Geiger-mode single-photon avalanche diodes (SPADs) have found widespread use over the past two decades \cite{itzler_dark_2014,jiang_ingaaspinp_2007,zhang_comprehensive_2009, Hillesheim}.

All single-photon detectors exhibit dark counts in the absence of light: in avalanche photodiodes, dark counts occur when avalanches are triggered by electrical carriers that are thermally generated or emitted by trapping levels in the semiconductor\cite{kang2003dark, spinelli_physics_1997}. The rate at which these avalanches are triggered, called the dark count rate, is heavily dependent on detector settings such as the quantum efficiency and dead time \cite{wen_origin_2018,zhang_comprehensive_2009,jiang_ingaaspinp_2007}. Dark counts are a source of noise in any application and increase with the bias voltage needed to enhance the quantum efficiency of the detector.  There have been several studies of dark current in avalanche photodiodes \cite{fishburn_fundamentals_2012,itzler_dark_2014} and numerical models have been developed to compare with experimental measurements \cite{sarbazi_statistical_2018}. 

However, a complete statistical characterization of the dark counts in 1550 nm SPADs has not been reported. The purpose of our research is to explore the dark counts of SPADs by using time-tagged photodetector measurements that can be used to generate histograms of their interarrival time distributions, probability distributions, and entropy rates. A simplifying assumption often made about the statistics of dark counts is that they follow a Poisson distribution. We examine this assumption and show that it is not generally valid without proper processing of the data. This processing is essential when one wants to accurately characterize the statistics of single photon emitters or novel light sources. 

It is long understood that Poisson statistics are not always an appropriate model for counts from photomultiplier tubes \cite{ralph_w_engstrom_photomultiplier_1980,saleh_photoelectron_1978}, image sensors such as CCD or CMOS sensors \cite{baer_model_2006}, general purpose digital pulse processing systems \cite{abbene_high-rate_2015}, and SPAD photocounts \cite{sarbazi_statistical_2018,gallivanoni_progress_2010}. However, it is still often assumed that their dark counts follow Poisson statistics \cite{kang2003dark} whose probability distribution is given by eqn. \ref{discretepoisson} \cite{tzou_method_2015,vinogradov_probability_2009}
\begin{equation}
\label{discretepoisson}
    P(n) =\dfrac{(rT)^ne^{(rT)}}{n!}, 
\end{equation}
where $P(n)$ is the probability of $n$ counts being detected in the time interval $T$, assuming $r$ is the average rate of detected counts. We will define $\lambda = rT$ as the average number of counts in the interval $T$. The detector quantum efficiency, $\eta$, relates the optical power to the observed photon count rate, $r$, as: 
\begin{equation}
\label{rate efficiency}
    r = \eta (P/h\nu),
\end{equation}
where $P$ is the average optical power incident on the detector, and $h\nu$ is the energy per photon.

In many experiments the dark count rates are relatively small in comparison with the photon count rates, therefore their effect on the distribution is negligible. However, there are other practical cases where the dark count rate is comparable to or only slightly less than the photon count rate. In these instances, we aim to quantitatively explore the dark count distributions and their influence on a weak attenuated coherent light source which is known to be Poisson \cite{mandel_optical_1995,loudon_quantum_2000}.

We study the deviations of dark current statistics from Poisson statistics due to dead time and afterpulsing \cite{saleh_photoelectron_1978,teich_1976} and propose a method for correcting the histograms. First, we recommend extending the detector dead time to eliminate the effects of afterpulsing as observed from the interarrival distributions. Second, we suggest calculating the adjusted occurrence rate of dark counts from the measured rate by using a standard formula. This enables us to determine the correct slope for the interarrival time distribution of a Poisson process. Lastly, we will show that a further step is necessary to correct the probability distributions for dead time which involves implementing an iterative algorithm \cite{mandel_inversion_1980,srinivas_deadtime}.

\section{Instrumentation}
\begin{figure}
    \centering
    \includegraphics[scale=0.5]{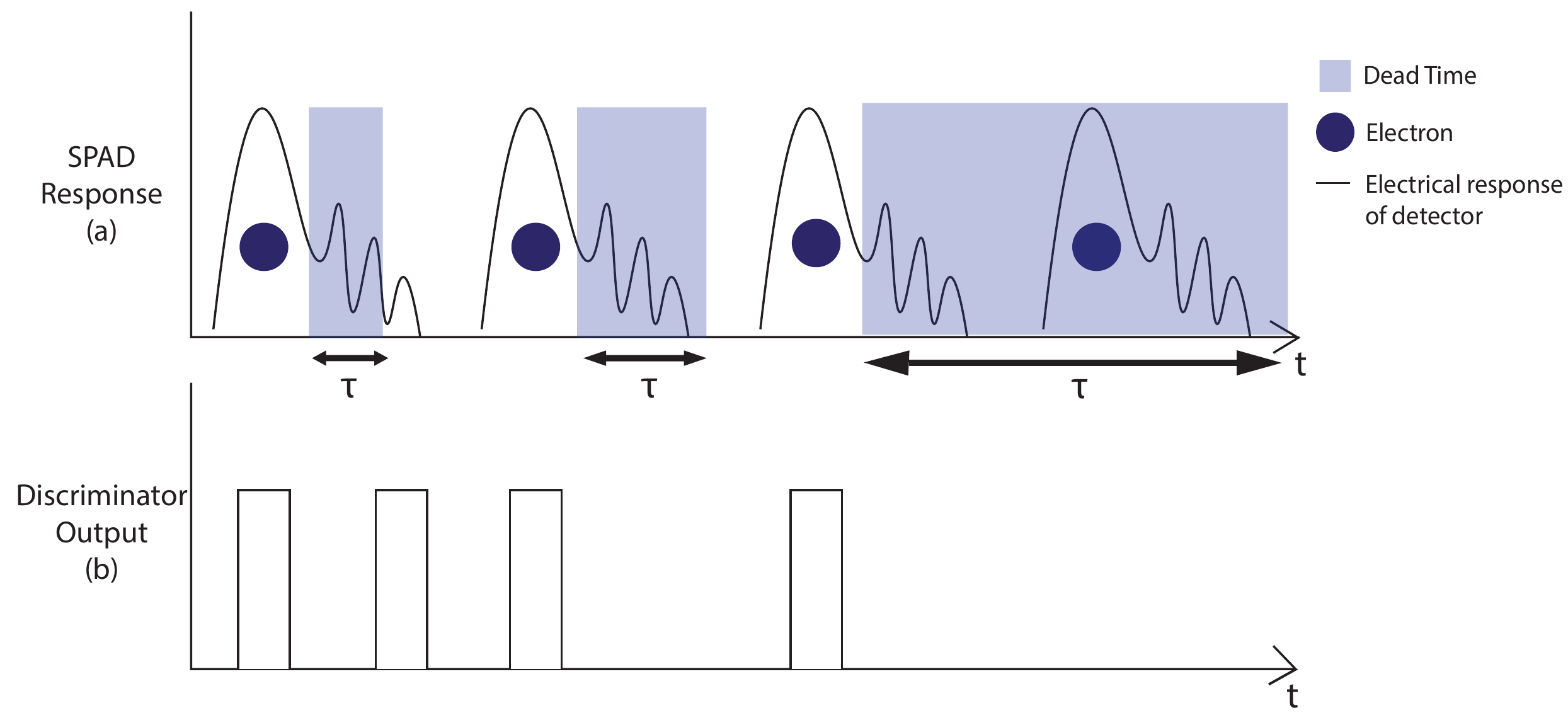}
    \caption{A schematic depicting the detector response to an electron carrier, which may be triggered by a photon, dark current, etc., in the presence of dead time. (a) The avalanche photodiode's electrical response (black curve) to a triggered electron (blue circle) including afterpulsing, the additional fluctuations seen after the initial response. After each response, there is a dead time (shaded blue boxes) when the detector cannot detect any incoming electrons. The dead time, $\tau$, is adjustable by the user. (b) Electrical signal sent from the output of the detector after the SPAD response passes through a discriminator. }
    \label{fig:Afterpulsing_schematic}
\end{figure}

Single-photon avalanche diodes are named as such because when the reverse bias of its p-n junction is raised above the breakdown voltage, just a single carrier can trigger an electrical avalanche process, leading to a measurable current \cite{spinelli_physics_1997,kang2003dark}. To detect a subsequent photon, the bias voltage must be reduced to near or below the breakdown value. Restoring the SPAD to its operative level, a process called quenching, is achieved by a quenching circuit that introduces a finite recovery time, known as dead time, during which the device cannot respond to another incident photon \cite{cova_avalanche_1996}. There are two main quenching modes: passive quenching (PQ) and active quenching (AQ). PQ SPADs are paralyzable detectors where photons arriving during the dead time are not counted and the dead time is extended \cite{sarbazi_statistical_2018}. Alternatively, AQ SPADs, a type of nonparalyzable detector used in our experiments, will not count photons arriving during the dead time nor will the dead time be extended by the quenching circuit.  If a carrier is triggered by photon absorption, the generated current will precisely mark the photon arrival time. However, avalanches can also be triggered by dark current, thereby marking the time of avalanche generation. Another phenomenon observed in SPADs is afterpulsing, a type of correlated noise found in real, non-ideal detectors where more than one electric pulse is generated per event due to traps holding extra charge carriers \cite{tzou_method_2015,ziarkash_comparative_2018,fishburn_fundamentals_2012}. Figure \ref{fig:Afterpulsing_schematic} depicts the detection behavior of AQ SPADs in the presence of dead time. Fig. \ref{fig:Afterpulsing_schematic}(a) shows the SPADs electrical response (black curve) to an electron (blue circle) which may be triggered by a photon or dark current. The additional fluctuations of the curve after the initial response are the afterpulsing. After each response, there is a pre-determined dead time, represented by the shaded blue boxes, during which the detector cannot detect any incoming electrons. In some detectors, the dead time $\tau$ is a parameter that can be controlled by the user, hence the shaded boxes are of different lengths. Fig. \ref{fig:Afterpulsing_schematic}(b) shows the electrical signal that is sent from the output of the detector after the SPAD response passes through a discriminator.

All our experimental data is taken using InGaAs Geiger-mode avalanche photodetectors from Aurea Technologies \cite{noauthor_nir_2019} operated in continuous mode and designed for 1550 nm wavelengths. Our SPADs have a range of selectable dead times from 1 $\mu$s to 999 $\mu$s and three quantum efficiency settings, 10\%, 20\%, and 30\%, which correspond to the percentage of incident photons that are detected and depend on the bias voltage applied to the APD. We observed dark current using two chosen dead time values: 20 $\mu$s and 500 $\mu$s. The first is the shortest dead time that can be used across all efficiency settings while in continuous mode and the second proved to be sufficiently long to remove the effects of afterpulsing.

\begin{figure}
  \centering
  \includegraphics[scale = 0.18]{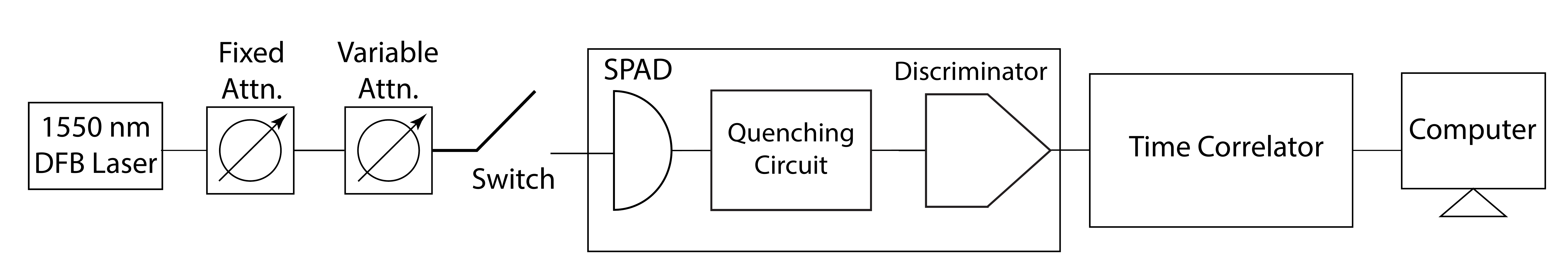}
\caption{Switch Open: Experimental setup used for measuring dark current. A SPAD set in continuous mode with no light entering it has its electrical output connected to the input of the time-tagging correlator. The time-tagged data is then sent to the computer to be post-processed. Switch Closed: Laser data experimental setup. A 1550nm distributed feedback (DFB) laser is connected through single-mode (SM) fiber to 110 dB of fixed attenuation and up to 30 dB of variable attenuation. Photons are then sent through SM fiber to SPAD. The electrical output of the SPAD is ultimately sent to the input of the time-correlator. The data is then sent to a computer to be post-processed.}
\label{fig:ExperimentalSetup}
\end{figure}

To experimentally measure dark current, detected events were recorded when no light was incident on the detector and the input was sealed and covered to minimize external light leakage (Fig. \ref{fig:ExperimentalSetup}- switch open). The precise arrival of events was determined using time-tagging measurements with a ChronoXea \cite{noauthor_time_2020}, a time correlator device from Aurea Technologies. This instrument accepts the electrical output of the detector and records the occurrence time with respect to an internal clock with a resolution of 13 ps. Afterward, the time-tagged data is post-processed on a computer. 

While using a dead time of  20 $\mu$s, around 500 dark counts per second were observed at 10\% efficiency; 5000 dark counts per second at 20\% efficiency, and 50,000 per second at 30\% efficiency. In this paper, we restrict ourselves to the analysis of dark counts at the 10\% and 20\% efficiency levels because at 30\% efficiency the dark current nears the saturation limit of the detector which makes it practical to use only while in gated mode.

%%%%%%%%%%   Experimental Setup Diagrams   %%%%%%%%%%%

\section{Interarrival distributions and adjusted occurrence rate}

\begin{figure}[h!]
  \centering
  \includegraphics[width=1\linewidth]{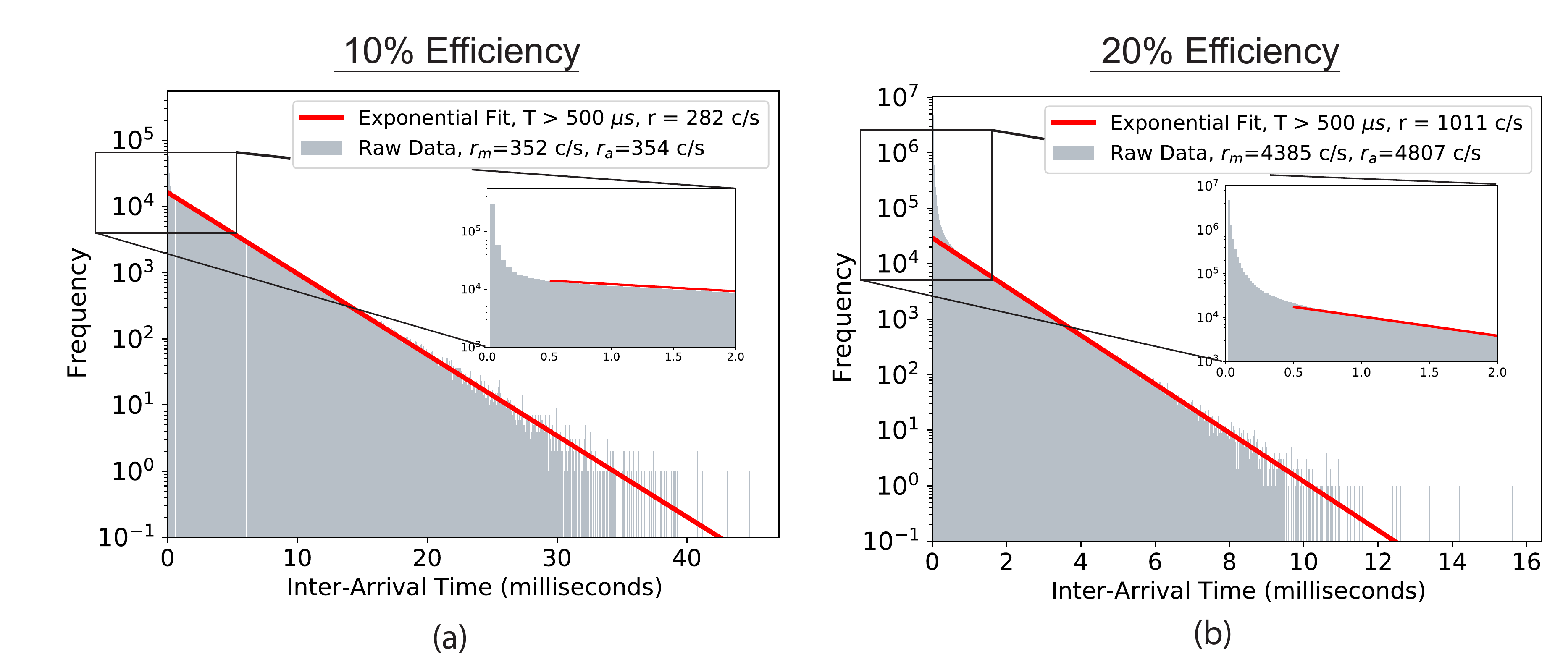}
\caption{Dark count data with SPAD settings of 20 $\mu$s dead time. (a) A histogram (grey) of the interarrival times between consecutive dark counts taken at 10\% efficiency. An exponential curve fit excluding the first 500 microseconds is shown in red. An inset image of the interarrival histogram with the first 2 milliseconds enlarged is included. (b) Interarrival time histogram at 20\% efficiency. }
\label{fig:20usInterarrival}
\end{figure}

The interarrival time distribution for a Poisson process is expected to be an exponential distribution shifted by the dead time \cite{wahl2011ultrafast}. Dead time hinders our ability to observe the true nature of the dark current distributions. One common way to account for dead time, $\tau$, is to calculate the adjusted occurrence rate, $r_a$, of counts if there were no dead time. This can be found using the experimentally measured occurrence rate, $r_m$, taken with dead time $\tau$ and the following equation \cite{larsen_simple_2009,lee_new_2000,neri_note_2010,lucke_counting}:
 \begin{equation}
 \label{truecps}
     r_a = r_m / (1 - (r_m\tau)).
 \end{equation} 
 This adjusted occurrence rate is often used to find fits for both the interarrival distributions and probability distributions. We calculated and plotted on a semi-log plot the interarrival times of the dark counts taken at 10\% and 20\% efficiency, both with a 20 $\mu$s dead time (Fig.~\ref{fig:20usInterarrival}(a-b)) and used eq. \ref{exponentialfit} to fit an exponential curve to the data and compare the slope of best fit, $r$, to the measured slope $r_m$ and adjusted slope $r_a$.

 \begin{equation}
 \label{exponentialfit}
     y(t) = A\exp[-r(t-\tau)]
 \end{equation}
In eq.\ref{exponentialfit}, $A$ is the frequency of counts in the first time bin, $r$ is the slope representing the occurrence rate of events per time $t$, and $\tau$ is the dead time.  The distribution for both data sets is exponential at longer interarrival times, but there is a large accumulation of dark counts present at very short interarrival times, seen in the inset figures of Fig. \ref{fig:20usInterarrival}(a) and Fig. \ref{fig:20usInterarrival}(b). This indicates the presence of afterpulsing which occurs on a time scale that is larger than the dead time used for these data sets. $r_a$ and $r_m$ are very similar in value because of the small dead time but are both much larger than the best fit, $r$. In the case of short dead times, the correction using eq. \ref{truecps} resulting in $r_a$ does not result in a better fit because the distribution is skewed by afterpulsing, not dead time. The only way to obtain the best fit is to remove the afterpulsing effect which can either be done through post-processing, such as analyzing only the bins after the first 500 microseconds as shown in the red curve of Fig.\ref{fig:20usInterarrival}(a-b), or experimentally by extending the dead time which we demonstrate next.

%%%%%%  20 us InterArrival%%%%%%

\begin{figure}[h!]
  \centering
  \includegraphics[width=1\linewidth]{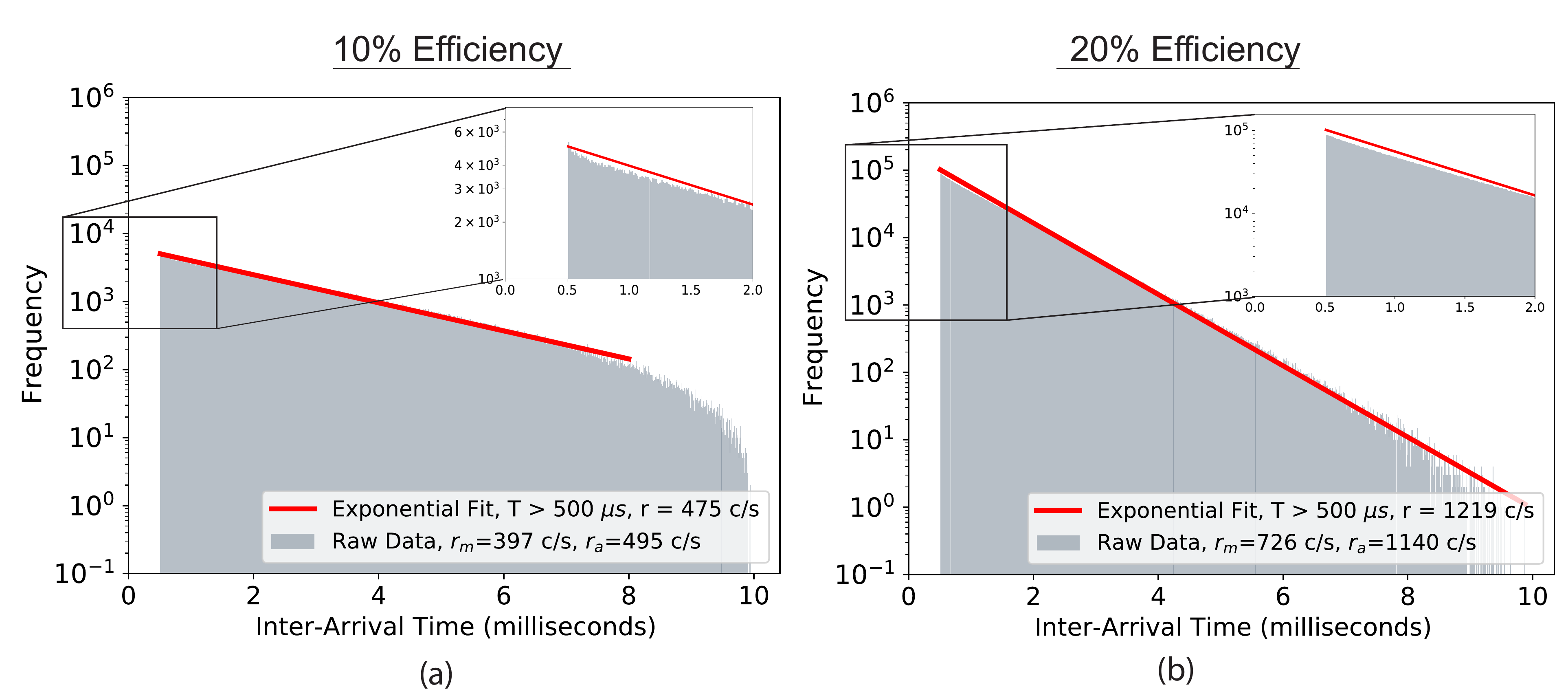}
\caption{Dark count data with SPAD settings of 500 $\mu$s dead time. (a) A histogram (grey) of the interarrival times between consecutive dark counts taken at 10\% efficiency. An exponential curve fit is shown in red. (b) Interarrival time histogram at 20\% efficiency. }
\label{fig:500usInterarrival}
\end{figure}

In order to experimentally eliminate the effect of afterpulsing, we increased our dead time to 500 $\mu$s which is well beyond the duration of the effect. The interarrival distributions for dark current at 10\% and 20\% quantum efficiency with a 500 $\mu$s dead time are shown in Fig.\ref{fig:500usInterarrival}(a-b). The afterpulsing effect is now successfully omitted in the interarrival distributions and they now closely resemble an exponential distribution. It should be noted that in Fig.\ref{fig:500usInterarrival}(a), the interarrival bins bend downwards in the last two milliseconds because they approach the maximum interarrival time measured by the Chronoxea, which in this case is 10 milliseconds. Therefore, the curve fit was only applied to data up to 8 milliseconds. The rate, $r$, determined by the curve fit in these figures is now very close to the calculated $r_a$ value. This demonstrates that eq. \ref{truecps} works better for fitting interarrival distributions with higher dead times.

%%%%  500 us InterArrival%%%%%%

\section{Probability distributions and iterative correction for dead time}
We have shown that afterpulsing distorts the interarrival distributions and that it is insufficient to simply ignore interarrival times smaller than 500 $\mu$s during post-processing. Instead, the electronic dead time must be extended in order to effectively suppress afterpulsing. In this regime of long dead time, using eq. \ref{truecps} to find $r_a$ provides a sufficient correction for dead time to fit the interarrival distribution of the data. However, next, we will demonstrate that this is not an effective correction to fit a Poisson distribution to the probability histogram, which expresses the probability of a given number of dark count occurrences within a fixed counting interval. Rather, a more rigorous method is required.

Finite dead time is an experimental feature in photon counting instrumentation that we would like to utilize and then correct by using a method of post-processing distributions. Mandel and Srinivas \cite{mandel_inversion_1980, srinivas_deadtime} offer a mathematical formula for the probability of photon counting in a counting interval which corrects for the effects of dead time in nonparalyzable single-photon detectors. It is an iterative method that allows for the counting probability of an ideal detector without dead time to be derived from the measured probability distributions observed with a real detector. Equation \ref{mandeleq} gives the probability, $p(n,T,0)$, that $n$ counts are registered in a time interval $T$ when the detector has no dead time in terms of probabilities that can be directly measured using a detector with dead time $\tau$. It assumes that the condition $T>n\tau$ is met throughout its use. When originally written, the algorithm required measurements for several different counting intervals ranging from $T$ to $T+n_0\tau$, where $n_0$ is the largest value of $n$ for which $p(n)$ is not negligible. However, with the common availability of time-tagging devices, one long measurement of time-tagged arrivals can be recorded. Afterwards, during post-processing, the arrivals can be grouped into counting intervals of duration $T+n\tau$ from which the probability $p(n,T+n\tau,\tau)$ can be estimated from the data, allowing us to solve for $p(n,T,0)$. We implement this more practical method rather than the multiple measurements suggested by Mandel. His equation is presented below:
\begin{equation}
\begin{split}
    p(n,T,0) =& p(n,T+n\tau,\tau)  \\ & +\sum_{r=0}^{n-1} [p(r,T+n\tau,\tau) - p(r,T+(n-1)\tau,\tau)], \quad n\geq 1, \\
    p(0,T,0) =& p(0,T,\tau).
    \end{split}
\label{mandeleq}
\end{equation}

\begin{figure}
  \centering
  \includegraphics[width=1\linewidth]{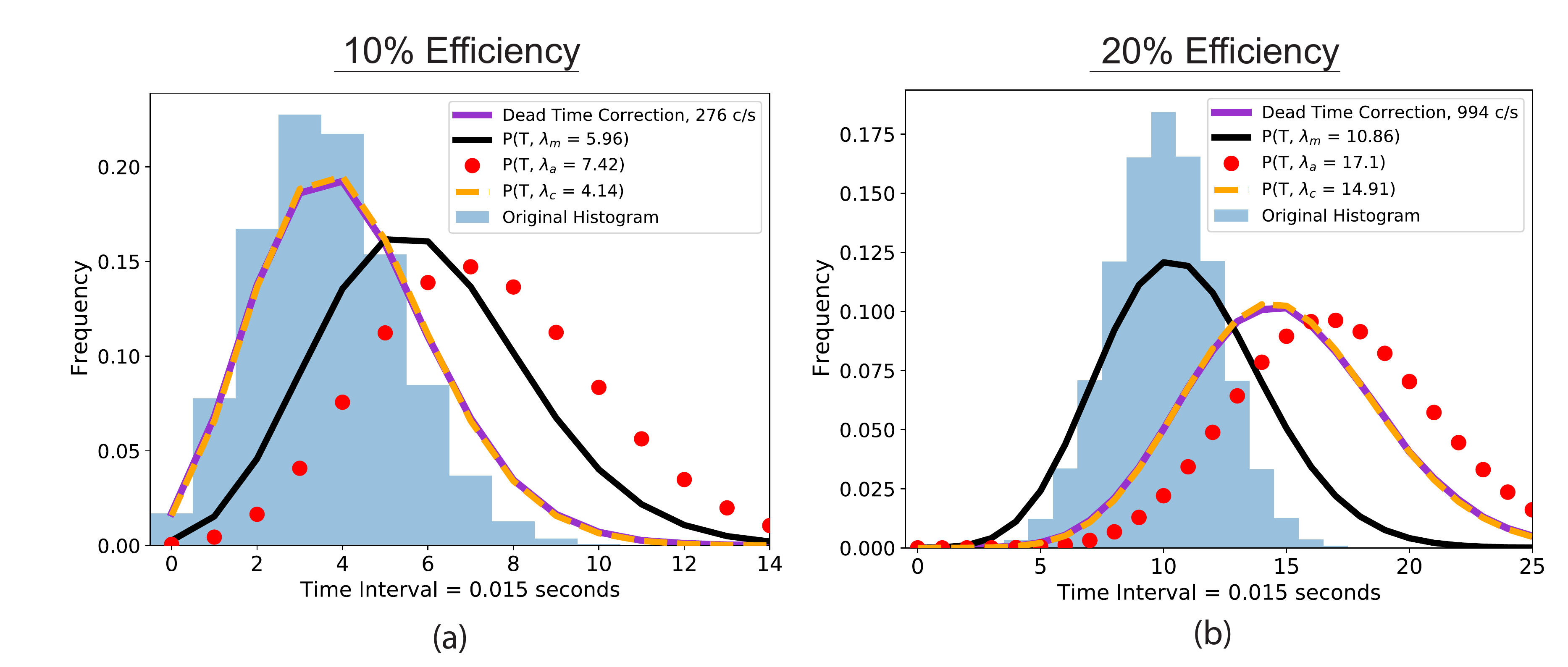}
\caption{Data with 500 $\mu$s dead time (a) A histogram of the normalized probability distribution (light blue) of dark counts which arrived in the time interval T = 0.015 seconds at 10\% efficiency  along with the dead-time corrected histogram (solid purple) generated from the iterative algorithm. Theoretical Poisson curves are included with varying values of $\lambda$, namely $\lambda_m$ (solid black), $\lambda_a$ (dotted red), and $\lambda_c$ (dashed orange). (b) Probability distribution for data taken at 20\% efficiency with the corresponding dead-time correction and several theoretical Poisson curves.}
\label{fig:500us_MandelT}
\end{figure}

This iterative algorithm was applied to our data resulting in dead-time corrected probability distributions. While an approximate algorithm (correct to third-order in $\tau /T$) for dead-time correction has previously been applied to photon counting data with photomultiplier tubes \cite{ODonnell:86}, to our knowledge, this work demonstrates the first comprehensive application of this exact algorithm to real SPAD data. Note that the Mandel algorithm corrects only for dead time, not afterpulsing, and that afterpulsing effects were removed by extending the dead time. Figs. \ref{fig:500us_MandelT}(a-b) depict the algorithm as applied to the 500 $\mu$s dead time data. Both figures include the original histogram (light blue) and the dead-time-corrected histogram (purple) plotted with a counting interval of T = 0.015 seconds. These histograms are compared to several different theoretical Poisson distributions, which for numerical stability and to avoid computing errors were calculated using:
\begin{equation}
    P(k,\lambda) = \exp[k\ln\lambda-\lambda-\ln\Gamma(k+1)],
\end{equation}
where $k = \lfloor\delta t/T\rfloor$ is the bin number with $\delta t$ as the interarrival time, $\Gamma$ is the Gamma function, and $\lambda = RT$ is the average number of counts with $R$ representing the occurrence rate of dark counts and $T$ representing the integration time. For all plots, three different Poisson distributions are plotted to compare to the original and corrected histograms. The first, shown in solid black, uses $\lambda_m = r_mT$ where $R = r_m$ is the measured occurrence rate. The second, shown in dotted red, uses $\lambda_a = r_aT$ where $R = r_a$ is the adjusted occurrence rate. The third, shown in dashed orange, uses $\lambda_c = r_cT$ where $R = r_c$ is the corrected occurrence rate calculated from the dead time corrected histogram.

%%%%%%%  500 us  Mandel %%%%%%%

Examining the probability distributions for data with detector settings of 500 $\mu$s (Figs.~\ref{fig:500us_MandelT}(a-b)), we see they are not closely aligned with the Poisson distributions using $\lambda_m$ or $\lambda_a$. The longer dead time affects the shape of the statistical distribution and simulates the effect of anti-bunching on the dark counts. The histograms are negatively skewed and sub-poissonian, meaning the statistical spread is narrower than a theoretical Poisson distribution of the same $\lambda$. 
 
 The long dead time makes this data set a good candidate for the iterative dead time correction algorithm and applying it appreciably shifts the histograms of the 10\% and 20\% efficiency data to more closely resemble the corresponding corrected Poisson distributions using $\lambda_c$. Here, the average number of counts lost due to the long dead time is high, allowing the Mandel algorithm to be effective. As described in section 2, the dark counts increase with increasing quantum efficiency. This means that the dark count rate is higher with the 20\% efficiency setting, and therefore more counts are lost due to the dead time than in the 10\% efficiency case. The dead time correction restores these lost counts, which is why the dead time corrected Poisson distribution has a higher count rate than the original histogram in the 20\% efficiency case. After applying the algorithm, the corrected distribution (purple) almost perfectly aligns with the Poisson distribution (orange) for the same $\lambda$ value in both cases. In the regime of long dead time, the effects of afterpulsing on the distributions are now negligible and the effects of dead time are dominant. Now, the Mandel algorithm can be used to correct for the effects of dead time. The iterative algorithm provides a correction to the actual histogram that can be used to derive a corrected $\lambda$ value that gives a perfectly fitting Poisson distribution. This correction method is more rigorous and much more effective than using eq. \ref{truecps} to adjust the $\lambda$ value.

\section{Dead time correction with attenuated laser source}
\begin{figure}
  \centering
  \includegraphics[width=1\linewidth]{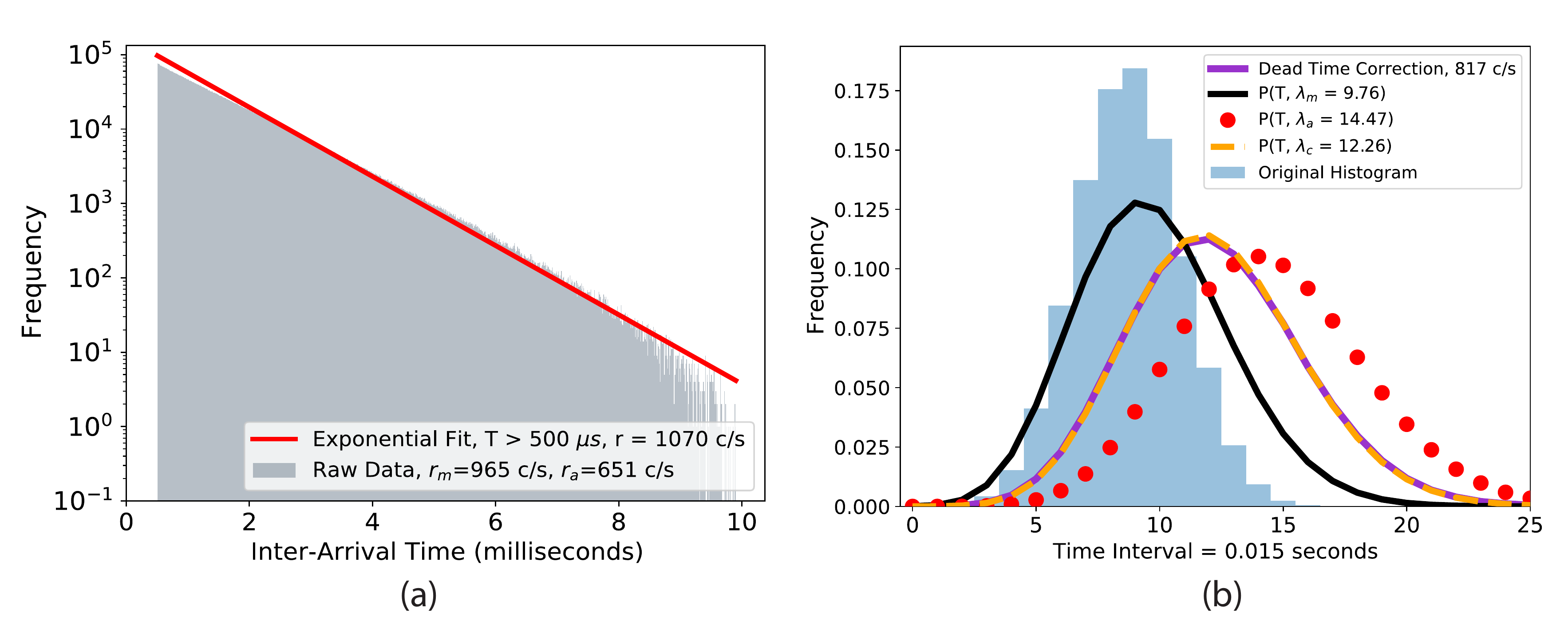}
\caption{Laser data (10 million points) with an average of 650 photons per second (adjusted for dead time) taken with detector settings of 10\% efficiency and 500 $\mu$s dead time. (a) Interarrival times on a log-linear plot (grey) with an exponential curve fit (red) based on the true photon counts detected per second, but excluding the first 500$\mu$s. (b)The original histogram (light blue) of the frequency of photon arrivals per time interval T. The dead time corrected histogram (purple) can be compared to the theoretical Poisson fit (dashed orange) for the corrected lambda value. Poisson fits are plotted using both the $\lambda_m$ (black) and $\lambda_a$ (dotted red)}
\label{fig:laserdata}
\end{figure}

Typically, SPADs are used to perform photon counting in systems using a light source. Below we analyze the photon distributions generated from a laser source and demonstrate that though the effects of dead time are still present, they can be adjusted for by using the dead time correction algorithm to find the corrected $\lambda$ value. In addition to the counts corresponding to the light source, dark counts occur as before. To produce light at the single photon level, a distributed feedback laser with a wavelength of 1550nm operated at 1 mW was attenuated by 140 dB using a combination of both fixed and variable attenuators connected by fiber. The detector was set to 10\% efficiency with a 500$\mu$s dead time which corresponds to approximately 225 dark counts per second. To simulate a scenario where the photon counts are on the same order of magnitude as the dark counts, we chose a laser attenuation level with a photon rate about twice as high as the dark counts, giving a total average of 525 measured counts per second. Just as in the experimental setup measuring dark counts, the electrical output of the SPAD was connected to the input of the time-tagging correlator, shown in Fig.\ref{fig:ExperimentalSetup}(switch closed), which recorded 10 million data points for each trial. The photon interarrival and probability distributions are shown in Fig. \ref{fig:laserdata}(a-b) along with the dead time corrected histogram and several Poisson plots.

%%%%%%%%  Laser Data  %%%%%%%%%%%%%
 Using this long dead time, the effects of afterpulsing are effectively eliminated (Fig. \ref{fig:laserdata}a). The probability distribution histogram (light blue) (Figs. \ref{fig:laserdata}b) again does not fit well with the Poisson distributions computed with $\lambda_m$ (black) and $\lambda_a$ (red) values and is sub-poissonian due to the long dead time. Applying the dead-time correction algorithm in Fig. \ref{fig:laserdata}(b) proves to correct for this poor fit and produces a corrected probability distribution (purple) that almost perfectly aligns with the Poisson distribution computed with its corresponding $\lambda_c$ value. This extension of the dead time past the regime of afterpulsing paired with the dead-time correction produces an accurate picture of the true distribution. It again demonstrates that a correction beyond adjusting the counts using eq. \ref{truecps} is necessary in order to generate a proper statistical description of the system data.

\section{Entropy rate of dark counts}

In the previous sections, we presented measurements of the dark counts in SPADs at 1550nm. All of these measurements consisted of histograms, which do not consider the temporal correlations or dynamical behavior of the measurements. In this section, we consider the  $\epsilon$-entropy rate of the SPAD dark counts, which relates to the entropy generation rate of a system as a function of its measurement resolution $\epsilon$. The $\epsilon$-entropy rate is a single metric that can simultaneously quantify both the shape of the interarrival time histogram and temporal correlations in the data. As we will show, the effects of the dead time and afterpulsing in SPAD dark count measurements are both captured using the $\epsilon$-entropy.

Considering the entropy rate as a function of the measurement resolution is useful for characterizing the information production of stochastic and deterministic processes in general \cite{gaspard1993noise,boffetta2002predictability,hagerstrom2015harvesting} and in particular, has important implications for random number generation based on physical entropy sources \cite{hart2017recommendations}. The detection of single photons is perhaps the best-known optical technique for generating entropy for random bit generation \cite{herrero2017quantum}.

In this section, we perform an analysis of the rate at which entropy can be harvested from the interarrival times of dark counts in our SPAD setup described above. In particular, we study how the entropy rate changes as a function of the interarrival time resolution $\Delta t$. While this is typically called the $\epsilon$-entropy, here we refer to it as the $\Delta t$-entropy to emphasize that we are studying the entropy production rate as a function of $\Delta t$. We will show that this single metric provides information about both the shape of the histogram as well as the temporal correlations present in the data.

The amount of information produced by a system per unit of time is given by its Shannon entropy rate. More precisely, the Shannon entropy rate is the average amount of new information obtained by measuring one additional sample with resolution $\epsilon$ given the history of all previous samples \cite{boffetta2002predictability}. Stochastic systems have an infinite Shannon entropy rate \cite{gaspard1993noise}.  However, information can only be extracted from the system at a finite rate, which is determined in part by the resolution of the measurement \cite{gaspard1993noise,boffetta2002predictability}. This motivated the definition of the so-called $\epsilon$-entropy by Shannon \cite{shannon1948mathematical}. The $\epsilon$-entropy has been studied for a variety of systems in theory and simulations \cite{gaspard1993noise,boffetta2002predictability} and experiments \cite{hagerstrom2015harvesting,hart2017recommendations}. Since here we are considering interarrival time measurements, for which the resolution is in terms of time, we denote the resolution of the interarrival time measurement as $\Delta t$.

In principle, one could compute the $\Delta t$-entropy directly from Shannon's definition $h_1(\Delta t)$ (see Eq. \ref{eq:informationentropy}). The first step is to generate a list of points in $d$-dimensional space using $d$ consecutive interarrival times. These vectors can be regarded as samples of a $d$-dimensional probability distribution. The entropy of this probability
distribution is sometimes referred to as the pattern entropy for patterns of length $d$. Computing the entropy in this way often requires an impractically large amount of data, especially as $d$ increases. 

Instead, we will use the correlation entropy rate $h_2(\Delta t;d)$, which we estimate using the Grassberger-Procaccia algorithm \cite{grassberger1983estimation}. The correlation entropy rate is a
lower bound on the information entropy rate, and in many cases is numerically very close to the
information entropy rate \cite{grassberger1983estimation}. A detailed discussion of the Shannon entropy rate and the correlation entropy rate and their analytic solutions for interarrival time measurements of Poisson processes can be found in the Appendix. For the SPAD system considered here, the only expected temporal correlations are due to afterpulsling, which should not create correlations between more than 2 or 3 consecutive events. Therefore, it is sufficient to consider only pattern lengths up to $d=3$.

%%%%%%%%%%%%%% ENTROPY FIGURE %%%%%%%%%%%%
\begin{figure}
\begin{subfigure}{.49\textwidth}
  \centering
  \includegraphics[width=.9\linewidth]{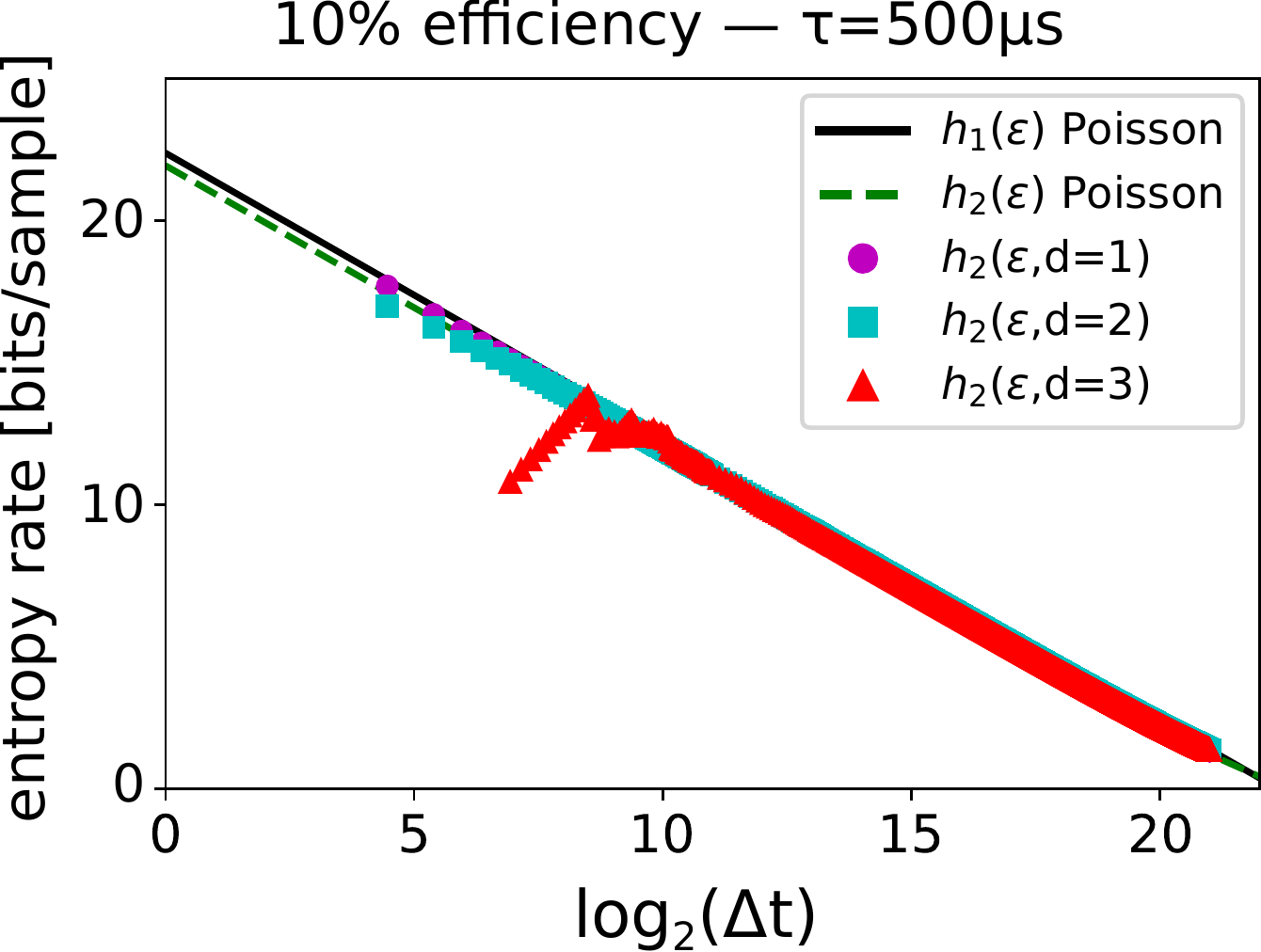}
  \caption{}
  \label{fig:entropy10eff500us}
\end{subfigure}
\begin{subfigure}{.49\textwidth}
  \centering
  \includegraphics[width=.9\linewidth]{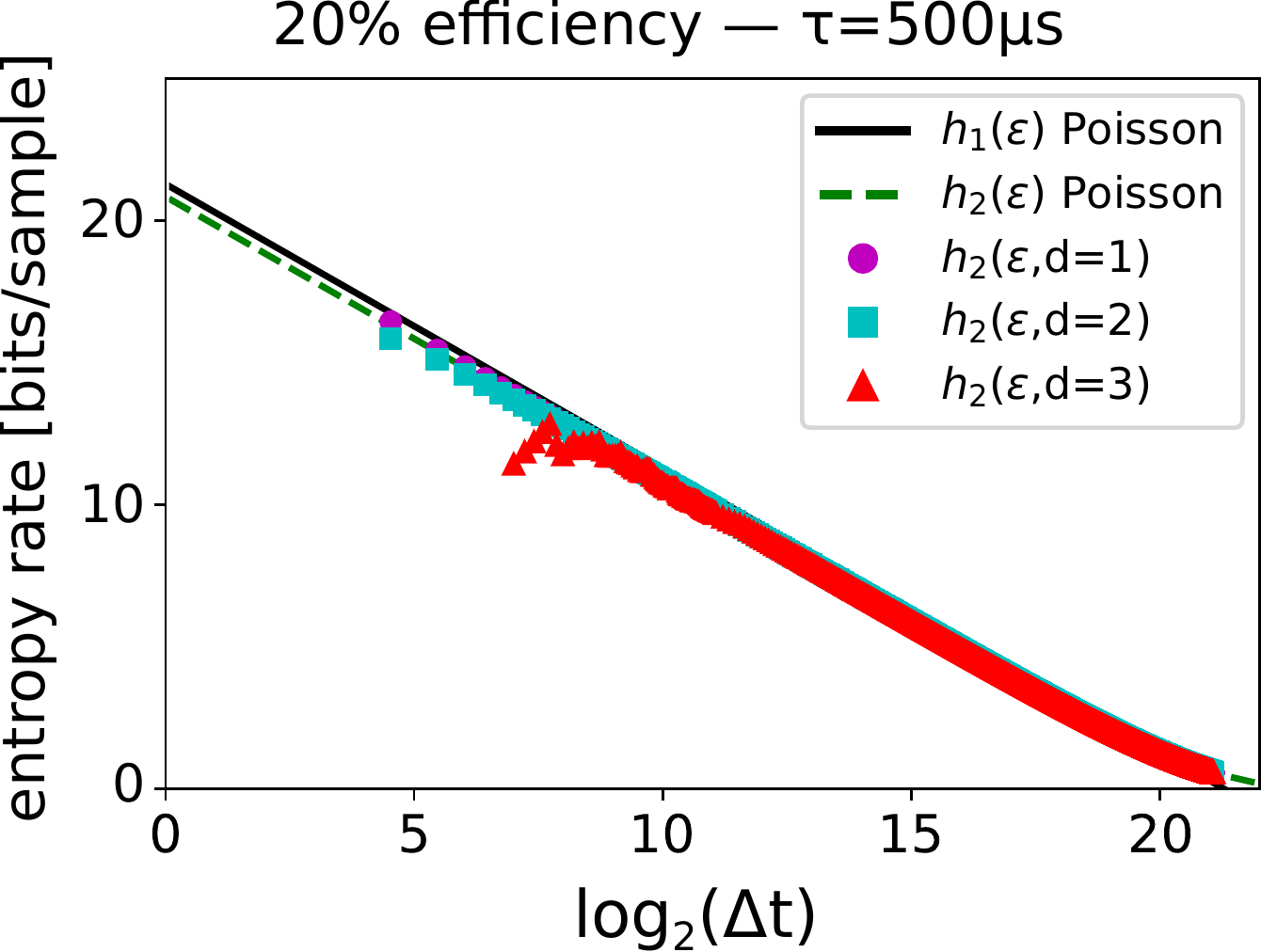}
  \caption{}
  \label{fig:entropy20eff500us}
\end{subfigure}
\caption{Entropy rate analysis for dark count measurements with 500 $\mu$s dead time and (a) 10\% quantum efficiency and (b) 20\% quantum efficiency. The analytic solution for $h_1$ (Eq. \ref{eq:PCentropy}) and $h_2$ (Eq. \ref{eq:PCrate2}) are shown as black and dotted green lines, respectively. The Grassberger-Procaccia estimates of $h_2(\Delta t,d)$ do not depend on $d$, implying that there are no intersample correlations. The sharp decrease as $\Delta t$ decreases for small $\Delta t$ in the $d=3$ curve (red) is due to the finite size of the dataset used. Further, the Grassberger-Procaccia $h_2$ estimates agree excellently with the analytic solution for $h_2$, suggesting that the dark count measurements are well-modeled by a Poisson process. All entropy rate estimates were computed using 2 million recorded interarrival times.}
\label{fig:entropy500us}
\end{figure}

%%%%%%%%%%%%%% ENTROPY FIGURE2 %%%%%%%%%%%%
\begin{figure}
\begin{subfigure}{.49\textwidth}
  \centering
  \includegraphics[width=.9\linewidth]{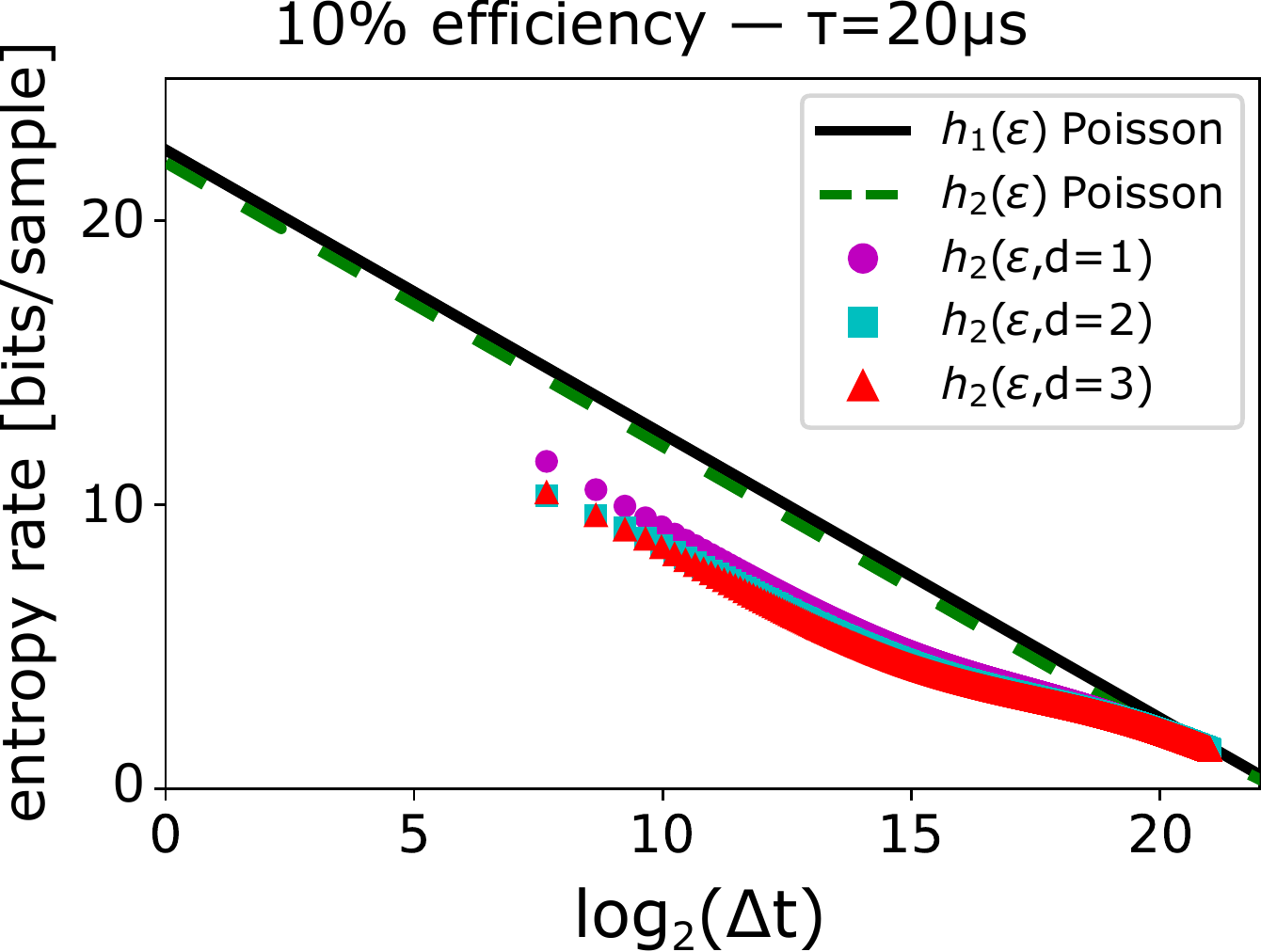}
  \caption{}
  \label{fig:entropy10eff20us}
\end{subfigure}
\begin{subfigure}{.49\textwidth}
  \centering
  \includegraphics[width=.9\linewidth]{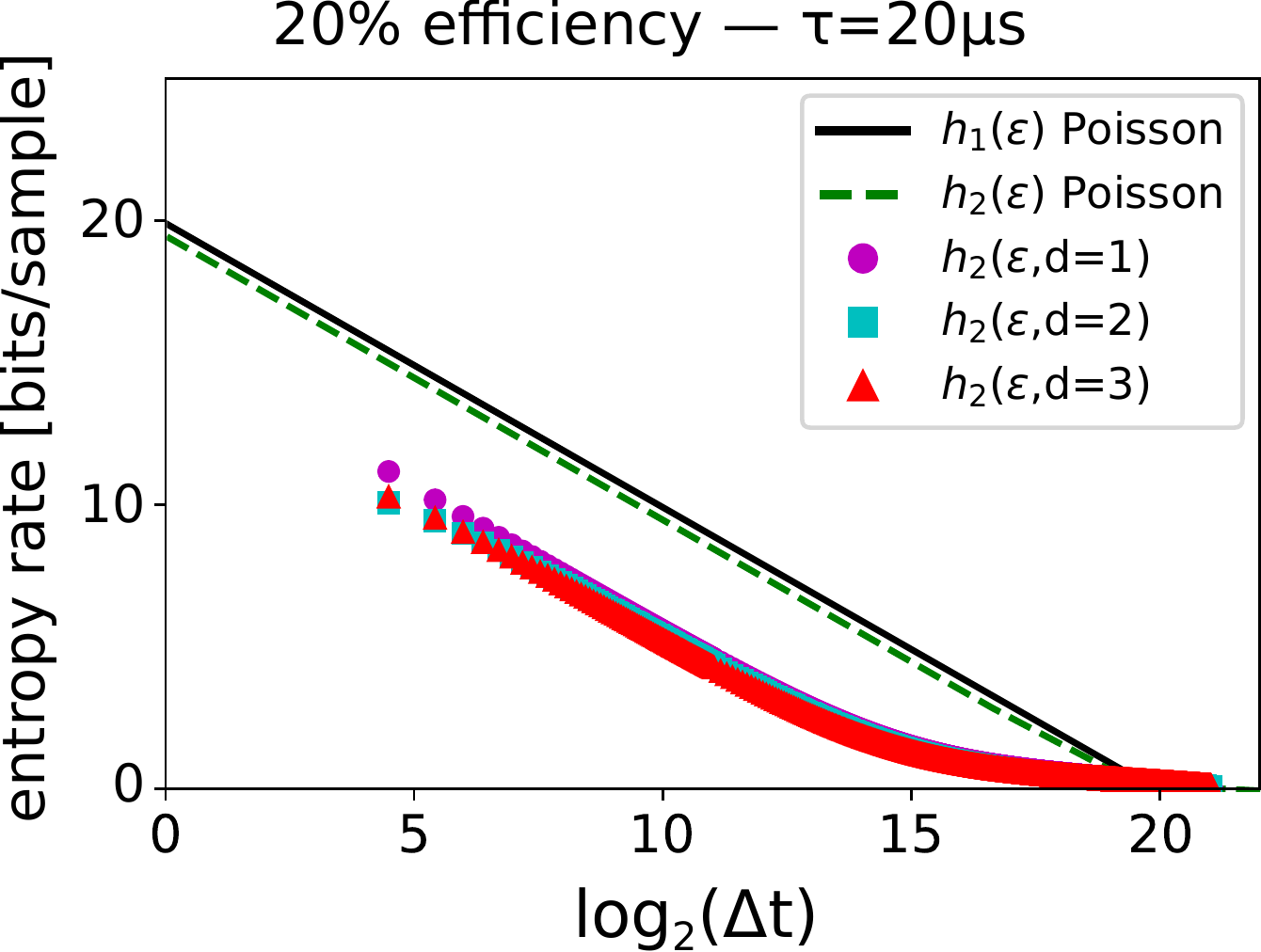}
  \caption{}
  \label{fig:entropy20eff20us}
\end{subfigure}
\caption{Entropy rate analysis for dark count measurements with 20 $\mu$s dead time and (a) 10\% quantum efficiency and (b) 20\% quantum efficiency. The analytic solution for $h_1$ (Eq.\ref{eq:PCentropy}) and $h_2$ (Eq.\ref{eq:PCrate2}) are shown as black and dotted green lines, respectively. The Grassberger-Procaccia estimates of $h_2(\Delta t,d)$ decrease with increasing $d$, implying that there are intersample correlations. The reduction with $d$ is stronger for the 20\% efficiency case, suggesting that there are stronger intersample correlations in this case, which is expected for afterpulsing. Further, for both efficiencies the Grassberger-Procaccia $h_2$ estimate severely underestimates the analytic solution for $h_2$ even for $d=1$, reflecting that the dark count interarrival time histogram is much more strongly peaked than the exponential distribution expected for a Poisson process. All entropy rate estimates were computed using 2 million recorded interarrival times.}
\label{fig:entropy20us}
\end{figure}

When the dead time is large, one expects there to be no intersample correlations since counts due to afterpulsing are eliminated. In this case, one might expect dark count measurements to be well-described by a Poisson process (Eq.\ref{eq:poissoncorrentropy}). Figure \ref{fig:entropy500us} shows that this is the case for the detectors used here for both the 10\% (Fig.~\ref{fig:entropy10eff500us}) and 20\% (Fig.~\ref{fig:entropy20eff500us}) quantum efficiency settings when the dead time is 500$\mu$s. The correlation entropy rate $h_2(\Delta t,d=1)$ estimated using the Grassberger-Procaccia algorithm agrees almost perfectly with the analytic result for the correlation entropy rate of a Poisson process with the same rate. This means that the distribution of interarrival times is extremely close to the exponential distribution expected for a Poisson process. Additionally, $h_2(\Delta t,d)$ is independent of $d$, as would be expected for a memoryless process. While essentially no intersample correlations are detected, $\lambda_d$ is also very low, so the entropy generation rate per unit time is quite low. Figure \ref{fig:entropy500us} also shows that, in this case, $h_2$ is an excellent approximation for $h_1$, as expected from Eqs.\ref{eq:PCentropy} and \ref{eq:poissoncorrentropy}.

For short enough dead times, on the other hand, one might expect to observe effects from afterpulsing. These effects might appear as intersample correlations as well as a probability distribution that is concentrated at lower interarrival times. Figure~\ref{fig:entropy20us} shows the Grassberger-Procaccia entropy rate estimate for dark count interarrival time measurements on our detectors with quantum efficiency 10\% (Fig.~\ref{fig:entropy10eff20us}) and 20\% (Fig.~\ref{fig:entropy20eff20us}). The striking disagreement between $h_2$ for a Poisson process (green dashed line) and the 1-$d$ Grassberger-Procaccia entropy rate (solid magenta line) estimate for both quantum efficiencies shows that these interarrival time distributions are significantly more peaked than the expected exponential distribution, as was shown in Figs. 3a-b. Further, the Grassberger-Procaccia entropy rate estimate decreases as $d$ increases from 1 to 2, indicating intersample correlations, which we attribute to afterpulsing.

\section{Conclusion}
Throughout our study, we find that both dead time and afterpulsing affect the probability distributions of dark counts and interarrival times and we quantitatively assess the assumption that dark current follows a Poisson distribution. We realize that the dead time of the detection circuitry must be adjusted to eliminate afterpulsing effects. With dead times that are non-negligible compared to the counting time intervals, a correction algorithm needs to be implemented. This algorithm \cite{mandel_inversion_1980, srinivas_deadtime} is iterative in nature and can be applied conveniently to time-tagged events measured by your detectors. As a result of these correction techniques, we are able to utilize SPADs for accurate determinations of the statistics of novel light sources in the 1550 nm range. 

\vspace{0.4cm}
\textbf{Acknowledgements:} N.M. and R.R. thank the Office of Naval Research Grant Number: N00142012139 (R.R.). We thank the experts at Aurea Technologies for their help with questions regarding the instrumentation. 

\vspace{0.4cm}
\textbf{Disclosures: }The authors declare no conflicts of interest.

\vspace{0.4cm}
\textbf{Data availability:}  Data underlying the results presented in this paper are not publicly available at this time but may be obtained from the authors upon reasonable request.

\section{Appendix: Entropy rates of time-tagged photon arrivals}

\renewcommand{\theequation}{\arabic{equation}}
In this Appendix, we define the information entropy rate and the Grassberger-Procaccia entropy rate, and we present the analytic solutions for these entropy rates for a Poisson process.

For a single variable system like the single photon interarrival time measurements considered here, a $d$-sequence is defined as $\mathbf{x}_d[n]=\{x[1],...,x[d]\}$. The information entropy $H_1$ can be defined as a function of time-tagging resolution $\Delta t$ and sequence length $d$ \cite{gaspard1993noise}:
\begin{equation}
\label{eq:informationentropy}
H_1(\Delta t,d)=-\sum_{k=0}^\infty p_k(\Delta t;d)\log_2(p_k(\Delta t;d))=\langle\log_2(p(\Delta t,d))\rangle,
\end{equation}
where $p_k(\Delta t;d)$ is the probability that a sequence of length $d$ is in the $k^{th}$ $d$-dimensional box of size $\Delta t$ and $\langle\rangle$ indicates the average over all boxes with non-zero probability. The information entropy rate per sample can be defined as a function of $\Delta t$ and $d$: $h_1(\Delta t;d)= [H_1(\Delta t;d) - H_1(\Delta t;d-1)]$; this is the additional amount of information obtained from the the next measurement given the previous $d-1$ measurements.

The probability $p_i(\Delta t;d)$ can be estimated from data using \cite{gaspard1993noise}

\begin{equation}
\label{eq:Pest}
    p_i(\Delta t,d) = \frac{1}{N_d}\{\mathrm{number\, of\, }d\mathrm{-sequences}\: \mathbf{x}_j \: \mathrm{s.t.} \, |\textbf{x}_i-\textbf{x}_j|<\Delta t\},
\end{equation}
where $N_d = N-d+1$ is the number of sequences of length $d$ in the data set, and $|\cdot|$ indicates a distance. This method of estimating the entropy rate has been used with some success for entropy rate generation for stochastic and/or chaotic systems \cite{gaspard1993noise,hagerstrom2015harvesting,hart2017recommendations,kawaguchi2021entropy}, but has the well-known problem that the estimation of $p_i(\Delta t;d)$ can require impractically large amounts of data when $\Delta t$ is small and $d$ is large.

The correlation entropy rate $h_2$, which is easier to estimate from experimental data, has been suggested as a substitute for the information entropy $h_1$ \cite{takens1983invariants,grassberger1983estimation}. The correlation entropy rate is a lower bound on the information entropy rate, and in many cases is numerically very close to the information entropy rate \cite{grassberger1983estimation}. Therefore, we use the correlation entropy rate $h_2$ as an estimate for $h_1$ (recognizing that $h_2\leq h_1$). The correlation $d-$block entropy is defined as:
\begin{equation}
\label{eq:correntropy}
H_2(\Delta t,d) = \log_2\bigg(\sum_ip_i^2(\Delta t;d)\bigg)=\log_2\langle p^2(\Delta t;d)\rangle,
\end{equation}
with $h_2(\Delta t;d)= [H_2(\Delta t;d) - H_2(\Delta t;d-1)]$ in units of bits per sample.

Equation \ref{eq:Pest} leads to $\sum p_i^2(\Delta t;d)=C(\Delta t;d)$, the well-known correlation sum \cite{grassberger1983estimation}. Equations \ref{eq:Pest} and \ref{eq:correntropy} are typically known as the Grassberger-Procaccia algorithm. We choose to use the infinity norm as was done by Takens \cite{takens1983invariants}. For each Grassberger-Procaccia entropy rate estimate, we use a data set of 2 million interarrival times.

For a white noise process, $h_2(\Delta t,d)$ increases linearly as the logarithm of $\Delta t$ decreases, and is independent of $d$ \cite{gaspard1993noise}.  On the other hand, for colored noise processes, $h_2(d)$ decreases as $d$ increases when $\Delta t$ is sufficiently small to resolve the time scale of the correlations \cite{gaspard1993noise}. The entropy rate additionally contains some information about the shape of the probability distribution: all else being equal, a broad distribution will have a greater entropy rate than a narrower distribution. This effect of the shape of the distribution is revealed explicitly by $h_2(\Delta t;d=1)$, which does not consider intersample correlations.

We now turn to the specific case of dark counts. It is often assumed that dark counts can be modeled by a Poisson process. The analytic solution for the information entropy generated per event $H_1(\Delta t;d)$ for a Poisson process is well-known to be:

\begin{equation}
\label{eq:PCentropy}
H(\Delta t)=\frac{(1-p_0)\log_2(1-p_0)}{p_0}+\log_2(p_0),
\end{equation}
where $p_0\equiv1-\exp{[-r_t\Delta t]}$. We note that this is independent of $d$ because Poisson processes are memory-less so there are no intersample correlations. The information entropy rate is then $h_1=r_m H_1$, where $r_m$ is the mean number of events measured per unit time.

Of course, single photon detectors have a dead time. For non-paralyzable detectors, the dead time $\tau$ does not affect the shape of the PDF; it only shifts it by $\tau$ \cite{wahl2011ultrafast,hart2017recommendations}. This does not affect $H$, the entropy per photon. The dead time does, however, affect the average rate of photons that are detected: $r_m=r_t/(1+r_t\tau)$ \cite{wahl2011ultrafast}. Thus the entropy rate becomes $h_1=r_mH$ in units of bits per unit time.

Further, if the probability of more than one event occurring in a single time bin is small ($r_t\Delta t<<1$), the information entropy rate for photon time-of-arrival measurements can be approximated as 
\begin{equation}
\label{eq:PCrate2}
h_1(\Delta t)=-r_m\log_2(\frac{r_t\Delta t}{e}).
\end{equation}

The correlation entropy per event for a Poisson process is independent of $d$ and can be computed directly from Eq. \ref{eq:correntropy} using $p_k=\exp{[-kr_t\Delta t]}(1-\exp{[-r_t\Delta t]})$:
\begin{equation}
\label{eq:poissoncorrentropy}
    H_2(\Delta t) = 2\log_2(1-e^{-r_t\Delta t})-\log_2(1-e^{-2r_t\Delta t}).
\end{equation}
In the limit that $r_t\Delta t<<1$, we can approximate the correlation entropy rate for a Poisson process with non-paralyzable dead time as

\begin{equation}
    h_2(\Delta t)=-r_m\log_2(\frac{r_t\Delta t}{2})
\end{equation}
in units of bits per second. We note that this is less than, but close to, $h_1$.

\bibliography{bib.bib}
\end{document}